\documentstyle[12pt]{article}

\renewcommand{\baselinestretch}{1.1}

\def\be{\begin{equation}}
\def\ee{\end{equation}}
\def\bea{\begin{eqnarray}}
\def\eea{\end{eqnarray}}
\def\beannmb{\begin{eqnarray*}}
\def\eeannmb{\end{eqnarray*}}

\def\no{\noindent}
\def\ol{\overline}
\def\ul{\underline}
\def\nnmb{\nonumber}
\def\rarrow{\rightarrow}

\def\mbm{{\rm \makebox[1em]{ }}}
\def\mbmm{{\rm \makebox[2em]{ }}}

\def\1{\={1}}
\def\2{\={2}}
\def\AA{$\mid$}
\def\MM{$_{\:}\rangle\mid$}
\def\NA{$\mid\! 0_{1}\rangle$}
\def\NB{$\mid\! 0_{2}\rangle$}
\def\NC{$\mid\! 0_{3}\rangle$}
\def\ND{$\mid\! 0_{4}\rangle$}
\def\NE{$\mid\! 0_{5}\rangle$}
\def\NF{$\mid\! 0_{6}\rangle$}

\def\ZZ{$\rangle$}
\def\ZA{\ZZ & \AA}

\def\BG{$\mid\,$}         

\def\gapA{\;\;\mid\!\!} 

\def\p1{ 1$\;$}
\def\m1{-1$\;$}
\def\pB{ 2$\;$}
\def\mB{-2$\;$}
\def\pC{ 3$\;$}
\def\mC{-3$\;$}
\def\pD{ 4$\;$}
\def\mD{-4$\;$}
\def\pE{ 5$\;$}
\def\mE{-5$\;$}
\def\pF{ 6$\;$}
\def\mF{-6$\;$}
\def\pG{ 7$\;$}
\def\mG{-7$\;$}
\def\pH{ 8$\;$}
\def\mH{-8$\;$}
\def\mwo{$\;$-1$\;$}
\def\mqrt{\!\!\!\!$-$\sqrt}
\def\sQ{$\:\sqrt{2}$}
\def\qQ{-$ \sqrt{2}$}
\def\sW{$\:\sqrt{8}$}
\def\qW{-$ \sqrt{8}$}

\begin{document}

\begin{titlepage}

\makebox{ }

\begin{tabbing}
xxxxxxxxxxxxxxxxxxxxxxxxxxxxxxxxxxxxxxxxxxxxxxxxxxxxxxxxxx\=xxxxxxxxxx\kill
 \> NUHEP-TH-00-81 \\ 
 \> SHEP-01-05 \\ 
\end{tabbing}
\vskip 0.4in

\begin{center}

{\large \bf 
$E_6$ unification model building III.\\ 
Clebsch-Gordan coefficients in $E_6$ tensor products \\
of the ${\bf 27}$ with higher dimensional representations
}

\vskip 0.4in

Gregory W.~Anderson 
                   $^{a}$ 
and
Tom\'{a}\v{s} Bla\v{z}ek$^{*}$ 
                   $^{b}$

\vskip 0.2in

$^a$
    {\em Department of Physics and Astronomy, Northwestern University,}\\
    {\em 2145 Sheridan Road, Evanston, IL 60208, USA}\\
\vskip 0.05in

$^b$
    {\em Department of Physics and Astronomy, University of Southampton,}\\
    {\em Highfield, Southampton SO17 1BJ, UK}\\
\vskip 0.05in

e-mail: {\em anderson@susy.phys.nwu.edu, blazek@soton.ac.uk}

\vskip 0.2in

January, 2001 \\

\end{center}

\vskip 0.7in

\begin{abstract}
$E_6$ is an attractive group for unification model building. However, the 
complexity of a rank 6 group makes it non-trivial 
to write down the structure of higher dimensional 
operators in an $E_6$ theory in terms of the states
labeled by quantum numbers of the Standard Model gauge group. 
In this paper, we show the results of our
computation of the Clebsch-Gordan coefficients
for the products of the {\bf 27} with 
irreducible representations of higher dimensionality:
${\bf 78}$, ${\bf 351}$, ${\bf 351^\prime}$, ${\bf \ol{351}}$, 
and ${\bf \ol{351^\prime}}$. Application of
these results to $E_6$ model building involving higher dimensional
operators is straightforward.

\end{abstract}

\vskip 0.5in


\vskip 0.1in

$^*${\footnotesize On leave of absence from 
the Dept. of Theoretical Physics, Comenius Univ., Bratislava, Slovakia.}

\end{titlepage}

\renewcommand{\thepage}{\arabic{page}}
\setcounter{page}{1}

%
%
%
%

\section{Introduction}

$E_6$ is the minimal simple gauge group which could accommodate
one family of the observed fermions, {\it and} a family of Higgs states, into a
single gauge multiplet.\cite{grs} Therefore, unification models based on $E_6$ 
can provide relationships for the measured charged fermion masses 
and quark mixing angles: thirteen unrelated independent parameters of 
the Standard Model of elementary particles, and at the same time 
a small set of $E_6$ symmetric operators may relate the charged 
fermion data both to the masses and mixings in the neutrino sector 
and to the parameters of the Higgs sector.
In this respect, $E_6$ provides a framework for the
most economic unified supersymmetric theories. 

As is well known the key feature among the observed masses of the three generations
of fermions is the inter-generational hierarchy. 
Any unified model has to explain the origin of the hierarchy in terms of the dynamics
of the underlying theory. In $E_6$ models, the hierarchy can follow from
the pattern of the symmetry breaking as the rank 6 group is broken down to
the Standard Model gauge group, possibly in a succession of steps.
The hierarchy may be explicitly realized in terms of higher dimensional 
operators containing the light states after the superheavy degrees of freedom 
are integrated out of the effective theory generated below the $E_6$ breaking 
scale $M_6$. 
\cite{Froggatt_Nielsen}
From the technical point of view, the construction 
of higher dimensional $E_6$ symmetric operators and their structure
in terms of the Standard Model states is a non-trivial task. 
Assuming that the light states occupy the fundamental 27-dimensional irreducible 
representation (irrep), a complete knowledge of the tensor products 
of the {\bf 27} irrep with larger irreps is required. 
For instance,
the $E_6$ symmetry allows for a higher dimensional operator containing 
the product of three ${\bf 27}$s and a ${\bf 78}$, suppressed by some
heavy scale $M_H\geq M_6$. If the ${\bf 78}$ acquires a vacuum expectation
value (vev) $v_6\approx M_6$ and all three ${\bf 27}$s contain light states, 
such an operator contributes to the generation of fermion mass matrices. In particular, for
the first two families it may generate entries suppressed by $v_6/M_H$. 
Yet, the predictivity of $E_6$ can only be utilized if the exact
form of the singlet in 
${\bf 27} \otimes {\bf 27} \otimes {\bf 27} \otimes {\bf 78}$ 
in terms of the Standard Model states is known. If two of the ${\bf 27}$s are 
contracted antisymmetrically, one needs to know the 
Clebsch-Gordan decomposition of the ${\bf 351}$ in the tensor product
${\bf 27}\otimes {\bf 78}$, then the decomposition 
of the ${\bf \ol{27}}$ in the product ${\bf 27}\otimes {\bf 351}$,
and, lastly, the decomposition of the singlet in the product 
${\bf 27}\otimes {\bf \ol{27}}$.
However, a complete information on general 
tensor products of exceptional group $E_6$ is difficult to obtain. \cite{pat_Verma} 
As for particular computations, to our knowledge 
only the Clebsch-Gordan decompositions of ${\bf 27}\otimes {\bf 27}$, 
${\bf 27}\otimes {\bf \ol{27}}$, and ${\bf 78}\otimes {\bf 78}$
are presently available in the literature. \cite{pat_E6,abI,abII}.
(We note in passing that a separate paper \cite{bky} obtains a subset 
of the results needed for the operator ${\bf 27^3 78}$    
by studying a branching chain of $E_6$. The results presented in this paper
are relevant for the case when the vev is acquired by a
zero weight state of the ${\bf 78}$.) 

In this paper, we continue in our earlier work \cite{abI,abII}  and
provide basic group-theoretical tools for a 
construction of higher-dimensional $E_6$ symmetric operators. 
In particular, we present the results of our computation of 
the Clebsch-Gordan coefficients (CGCs) for the tensor products
involving 27, 78, and 351-dimensional representations, the lowest 
dimensional irreps in $E_6$.
%
%
Section 2 contains some necessary mathematical background for our study,
mostly concerned with lowering in the weight system of these irreps.
Our main results can be found in section 3, where we also comment 
on the construction and properties of the weight systems of the resulting 
irreps. Section 4 contains the summary, while one appendix provides details
on the lowering relations in the presence of degenerate weights.

\section{Mathematical Preliminaries}
\label{sc:MathPrelim}

In this work we consider tensor products of the fundamental 27-dimensional irrep
with higher dimensional 78 and 351-dimensional representations of $E_6$.
In particular, we compute the Clebsch-Gordan coefficients for the products\\
\parbox{12.7cm}
{\begin{eqnarray*}
{\bf 27}\otimes {\bf \:   \;\;78\:      } &=& {\bf     1728 } \oplus  {\bf \;\; 351 } \oplus  
                                            {\bf   \;  27 },                                 \\
{\bf 27}\otimes {\bf \:    351\:        } &=& {\bf \ol{7371}} \oplus  {\bf \ol{1728}} \oplus  
                                            {\bf \ol{ 351}} \oplus  {\bf \ol{  27}},         \\ 
{\bf 27}\otimes {\bf \:    351^{\prime} } &=& {\bf \ol{7722}} \oplus  {\bf \ol{1728}} \oplus  
                                            {\bf \;\ol{27}}                        ,         \\
{\bf 27}\otimes {\bf \:\ol{351}\:       } &=& {\bf \ol{5824}} \oplus  {\bf     2925 } \oplus  
                                            {\bf      650 } \oplus  {\bf       78 },         \\ 
{\bf 27}\otimes {\bf \:\ol{351^{\prime}}} &=& {\bf     3003 } \oplus  {\bf \ol{5824}} \oplus  
                                            {\bf      650 }.                                 
\end{eqnarray*}}
\hfill 
\parbox{0.7cm}
{\begin{eqnarray*}
   (\arabic{equation}{\mbox{a}}) \\ 
   (\arabic{equation}{\mbox{b}}) \\ 
   (\arabic{equation}{\mbox{c}}) \\ 
   (\arabic{equation}{\mbox{d}}) \\ 
   (\arabic{equation}{\mbox{e}}) 
                   \label{eq:dims_in_products}
\end{eqnarray*}}\\ 
\stepcounter{equation}
Before we discuss the construction of the weight systems of the irreps 
on the right hand side of these relations let us start first with 
the rules for the construction of the irreps on the left.

The key ingredient of our procedure is the lowering operation which is used to construct
a complete weight system by successive application of generators $E_{-\alpha_1},\ldots E_{-\alpha_6}$.
These are the generators which lie
outside of the diagonal Cartan subalgebra and correspond to the six simple roots of $E_6$. 
(Our choice of generators is described in more detail in \cite{abII}, 
and basically follows the standard conventions of \cite{Slansky_PhysRep}.)
The six generators act as ladder operators --- 
at each level the weight of the new state is obtained from the weight at the previous level
by subtracting (in the weight space) the respective simple root:
\be
   E_{-\alpha_i}|\, w\rangle\, = N_{-\alpha_i,\,w}|\, w-\alpha_i\rangle.
\label{eq:dfnN}
\ee
For the weight systems of the ${\bf 27}$ and ${\bf 78}$ constants $N_{-\alpha_i,w}$ satisfy
(see eq.(12) in \cite{abII})
\be
|N_{-\alpha_i,(w)_j}|^2 =          \langle \alpha_i| \,w\,\rangle\, 
                           +\, |\, \langle (w)_j\, |\, (w)_i\rangle|^2\;\; 
                               |N_{-\alpha_i,\,w+\alpha_i}|^2.
\label{eq:N-}
\ee
It is understood that the new state $|\, w-\alpha_i\rangle$ does not exist
if $N_{-\alpha_i,w}=0$  or the r.h.s. of (\ref{eq:N-}) turns out to be negative. 
The subscript on the weight $(w)$
%
is only relevant for the six degenerate zero weight states
of the ${\bf 78}$ and is to be ignored for non-degenerate weights. 
In fact, the second term on the r.h.s. of (\ref{eq:N-}) never contributes when one constructs
the weight system of the ${\bf 27}$ as there are no higher multiplets than doublets
for any $SU(2)$ subgroup. 

We remark that throughout this work, and consistent with our previous studies \cite{abI,abII}, 
the lowering phase convention which always fixes constants $N$ to be real and non-negative
\be
    N_{-\alpha_i,\, w} \ge 0
\label{eq:Npositive}
\ee
is adopted for any simple root $\alpha_i$ and any weight system. Then
for the zero weight states of the ${\bf 78}$ the inner product in (\ref{eq:N-}) can
be expressed as
\be
\langle (w)_i\, |\, (w)_j\rangle = |A_{ij}| / 2,
\label{eq:ww_in_78}
\ee
where $A_{ij}\equiv \langle \alpha_i\, |\, \alpha_j\rangle$ are the elements 
of the Cartan matrix of $E_6$. \cite{Slansky_PhysRep,abII} 
This result follows from the decomposition of the ${\bf 78}$ weight states 
into the states of the fundamental representations in the tensor product 
${\bf 27}\otimes {\bf \ol{27}}$. \cite{abI}

In the appendix, we derive a generalized relation for $N_{-\alpha_i,w}$ for a weight system 
with multiple degenerate weights at different levels. We now discuss how to apply general 
formula (\ref{eq:app3}) 
to the weight systems of the 351-dimensional representations which appear on the left in 
eq.(\ref{eq:dims_in_products}b--e). These irreps, although already rather large, are still special 
because for each weight subspace a basis can be defined such that the application 
of a lowering ladder operator results in a single basis state, as indicated in eq.(\ref{eq:dfnN}). 
(For larger irreps, there are lowerings which lead to a linear combination 
of the basis states regardless of the basis definition.)
Moreover, if weight $(w)$ is degenerate and a state with weight $(w-\alpha)$ exists,
then the $(w+\alpha)$ weight state is either non-degenerate or does not exist at all.
Thus for the weight system of the ${\bf 351^\prime}$ or ${\bf 351}$ 
eq.(\ref{eq:app3}) reduces to a simple form
\be
|N_{-\alpha,\,(w)_a\rarrow (w-\alpha)_{A_a}}|^2 =  \langle \alpha|\, w\rangle\, 
                                        +\, | \langle (w)_a\, |\, (w)_c\rangle|^2\;\; 
                                          |N_{-\alpha,\, (w+\alpha)\rarrow (w)_c}|^2,
\label{eq:N-deg}
\ee
where, formally, the summation over $c$ is assumed in the last term, but no more than one state actually
contributes. Concrete applications of this formula are provided at the end of the section.

Compared to eq.(\ref{eq:N-}) both weights $(w)$ and $(w-\alpha)$ can now be degenerate. 
We find, however, that in the ${\bf 351^\prime}$ or ${\bf 351}$ the $(w+\alpha)$
weight state does not exist if $(w-\alpha)$ is a degenerate weight.
Hence if both $(w)$ and $(w-\alpha)$ are degenerate, 
$A_a$ can be set to $a$ by definition and (\ref{eq:N-deg}) can be
simplified even further:
\be
|N_{-\alpha,\,(w)_a\rarrow (w-\alpha)_{a}}|^2 =  \langle \alpha|\, w\rangle, 
\qquad\mbox{\rm for both }\:(w)\:\: \mbox{\rm and }\:\:(w-\alpha)\:\: \mbox{\rm degenerate.}
\label{eq:N-deg2}
\ee
This shows that the definition of the basis states (and their subscript labeling) 
in the degenerate subspaces of the ${\bf 351^\prime}$ or ${\bf 351}$ 
can be induced from the basis states at the previous level.
However, once $(w)$ is found degenerate, how do we know if $(w-\alpha)$
is going to be degenerate and what the dimensionality of this subspace is going to be?
Similarly to the  case of the ${\bf 78}$ weight system \cite{abII} the decomposition of the 
351-dimensional irreps into the states of the fundamental representations can be recalled.
The product 
${\bf \ol{27}}\otimes {\bf \ol{27}} = {\bf 351^\prime} \oplus  {\bf 351} \oplus  {\bf 27 }$
is conjugated to the product studied in ref.\cite{pat_E6}.
We refer to this work to claim that all degenerate weight subspaces in the 
${\bf 351^\prime}$ (or ${\bf 351}$) are of the same dimensionality and that
the degenerate weights follow the weight system of the ${\bf 27}$.
In the end it thus turns out that 
complete bases in the degenerate weight subspaces of the ${\bf 351^\prime}$ 
can be obtained
%
%
starting from the four $(100000)$ weight states at level 8 of the ${\bf 351^\prime}$
\cite{footnote_1} :\\
\parbox{12cm}
{
  \beannmb
      |\,(100000)_3\rangle &=& E_{-\alpha_3} \: |\,(1\ol{1}2\ol{1}0\ol{1})\rangle \,/\sqrt{2},\\
      |\,(100000)_4\rangle &=& E_{-\alpha_4} \: |\,(10\ol{1}2\ol{1}0)\rangle      \,/\sqrt{2},     \\
      |\,(100000)_5\rangle &=& E_{-\alpha_5} \: |\,(100\ol{1}20)\rangle \,/\sqrt{2},     \\
      |\,(100000)_6\rangle &=& E_{-\alpha_6} \: |\,(10\ol{1}002)\rangle     \,/\sqrt{2}.      
  \eeannmb
}
\hfill
\parbox{3cm}
{
  \bea
    { }& &{ }\nnmb\\
    { }& &{ }      \label{eq:100000in351P}  \\
    { }& &{ }\nnmb
  \eea
}\\
With lowering convention (\ref{eq:Npositive}) 
the remaining 26 degenerate weight subspaces at lower levels can be specified as
$\,
|\,(w)_a\rangle = E_{-\alpha_A}\ldots E_{-\alpha_B}$ 
$|\,(100000)_a\rangle
$, 
$
a=3,4,5,6, 
$
%
%
where $E_{-\alpha_A}\ldots E_{-\alpha_B}$ is the lowering path 
leading to state $|(w)\,\rangle $ in the ${\bf 27}$.
For the ${\bf 351}$ the only difference is that the degenerate weight subspaces are five-dimensional
and the relations analogous to (\ref{eq:100000in351P}) also include 
\be
   |\,(100000)_2\rangle = E_{-\alpha_2} \:  |\,(02\ol{1}000)\rangle \,/\sqrt{2} 
   \label{eq:100000in351}        
\ee
when computing the $(100000)$ states at level 7 of this irrep.
We note that with this notation 
%
%
the inner product in any degenerate weight subspace of both
the ${\bf 351^\prime}$ and ${\bf 351}$ satisfies 
\be
   \langle (w)_a\, |\, (w)_b\rangle = |A_{ab}| / 2
\label{eq:ww_in_351s}
\ee
(where $a,b = 3,4,5,6$ for the ${\bf 351^\prime}$, and 
       $a,b = 2,3,4,5,6$ for the ${\bf 351}$), 
in a close similarity to the degenerate zero weight 
subspace of the ${\bf 78}$, eq.(\ref{eq:ww_in_78}). 

As an example of the application of formula (\ref{eq:N-deg}) consider all possible
lowerings of the state $|\,F_5\rangle$ in the ${\bf 351^\prime}$ where, for brevity, 
$F$ stands for the $(100000)$ weight. Three different states at the next level can be obtained:
$  E_{-\alpha_1}\, |\,F_5\rangle = N_1\: |\,(\ol{1}10000)_5\rangle $,
$  E_{-\alpha_4}\, |\,F_5\rangle = N_4\: |\,101\ol{2}10\rangle $, and
$  E_{-\alpha_5}\, |\,F_5\rangle = N_5\: |\,1001\ol{2}0\rangle$.
Based on (\ref{eq:N-deg}) the constants are
\[
 \begin{array}{ccccccc}
  |N_1|^2 &=& 1 &+&                              0  &=& 1          \, , \\
  |N_4|^2 &=& 0 &+& (\frac{1}{2})^2   (\sqrt{2})^2  &=& \frac{1}{2}\, , \\  
  |N_5|^2 &=& 0 &+&             1^2 \;(\sqrt{2})^2  &=& 2          \, ,
 \end{array}
\]
as we have already showed in the second and third equation (\ref{eq:100000in351P}) that 
\[
  N_{-\alpha_4,\,(F+\alpha_4)\rightarrow (F)_4} \, = \,
  N_{-\alpha_5,\,(F+\alpha_5)\rightarrow (F)_5} \, = \, \sqrt{2}.
\nonumber
\]
Implicitly, we also used the fact that (\ref{eq:100000in351P}) 
represents the only way how the $(100000)$ weight states can be obtained
from the states at the previous level.
Note that (\ref{eq:N-deg2}) could be used to compute $N_1$ since both $(100000)$
and $(\ol{1}10000)$ are degenerate weights.

Finally, we remark that the properties of the ${\bf \ol{351^\prime}}$ and ${\bf \ol{351}}$ are easily
derived from the properties of the ${\bf 351^\prime}$ and ${\bf 351}$ after the Dynkin coordinates 
[and any other indices in Dynkin formalism, like e.g., the labeling of states in 
eq.(\ref{eq:100000in351P})]
$1$ and $2$ are exchanged with $5$ and $4$, respectively.

\section{Construction of Clebsch-Gordan Coefficients}

Tensor products in eq.(\ref{eq:dims_in_products})
can be expressed in terms of the highest weights as\\
\parbox{14cm}
{\begin{eqnarray*}
(100000)\otimes(000001) &=& (100001)\oplus (000100) \oplus (100000)                , \\ 
(100000)\otimes(000100) &=& (100100)\oplus (000011) \oplus (010000) \oplus (000010), \\ 
(100000)\otimes(000020) &=& (100020)\oplus (000011)                 \oplus (000010), \\     
(100000)\otimes(010000) &=& (110000)\oplus (001000) \oplus (100010) \oplus (000001), \\ 
(100000)\otimes(200000) &=& (300000)\oplus (110000) \oplus (100010)                . 
\end{eqnarray*}}
\hfill 
\parbox{1cm}
{\begin{eqnarray*}
   (\arabic{equation}{\mbox{a}}) \\ 
   (\arabic{equation}{\mbox{b}}) \\ 
   (\arabic{equation}{\mbox{c}}) \\ 
   (\arabic{equation}{\mbox{d}}) \\ 
   (\arabic{equation}{\mbox{e}}) 
                   \label{eq:weights_in_products}
\end{eqnarray*}}\\ 
\stepcounter{equation}
For each product we start with the construction of the weight system of the first irrep on 
the r.h.s..
The highest weight state of this irrep is non-degenerate and can always be expressed as 
a trivial combination of the highest weight states of the two irreps on the left-hand side,
with the CGC being equal to $+1$. 
In the absence of a simple method how to determine the bases in the degenerate weight subspaces
which follow at lower levels for each of these irreps, we compute directly the complete weight
system in each case. However, note that simple lowering (\ref{eq:dfnN}) does not necessarily
hold for weights with multiple degeneracies, as is discussed in the appendix. States at lower 
levels are then computed by successive lowerings 
applied to the states of the 
${\bf 27}$, and ${\bf 78}$ in case $(a)$ or 
one of the 351-dimensional irreps in cases $(b)$--$(e)$. These lowerings were described in detail
in the previous section. The computed state is accepted and kept as a new basis state if it
cannot be expressed as a linear combination of the previously obtained basis states with the same weight.

It is not necessary to show the Clebsch-Gordan coefficients for every linearly independent state, 
since there are many states with the same CGCs. Instead, we
present the results just for the dominant weight states. Dominant weights are weights
with all Dynkin coordinates non-negative. The CGCs for the remaining states can then be 
determined using the charge conjugation operators \cite{pat_CCO, footnote_2}, or in a straightforward way by
direct lowering.
In tables \ref{t:paths_1728}--\ref{t:paths_3003} we present lowering paths for the
dominant weight states of the ${\bf 1728}$, ${\bf \ol{7371}}$, 
${\bf \ol{7722}}$, ${\bf \ol{5824}}$, and ${\bf 3003}$ irreps. In our abbreviated notation,
lowering path, let's say, 3421 stands for $E_{-\alpha_3} E_{-\alpha_4} E_{-\alpha_2} E_{-\alpha_1}$
applied (from the right) to the highest weight state.  
The lowering paths in tables \ref{t:paths_1728}--\ref{t:paths_3003} actually specify 
our choice of bases for particular dominant weight subspaces.
Explicit Clebsch-Gordan decomposition of the dominant weight states 
is important because, typically, the multiplicity of degeneracy (i.e., the dimensionality
of the weight subspace)
changes compared to the degeneracy at the previous level. Clearly, that is why these states cannot be 
obtained by generalized charge conjugation from the states at the previous levels.
Moreover, it is important to check the completeness of a reducible dominant weight subspace.
If it is impossible to complete its basis by lowering the states at the previous level, 
new weight systems open up and the remaining basis vectors are their highest weight states.
This is what happens for every dominant weight in the tensor products 
${\bf 78}\otimes {\bf 78}$ \cite{abII}, ${\bf 27}\otimes {\bf \ol{27}}$ \cite{abI},
or ${\bf 27}\otimes {\bf 27}$ \cite{abI} studied in the earlier work.
However, this property of the dominant weights is no longer true for the products studied here.
We now discuss shortly the dominant weights in each of the products in (\ref{eq:weights_in_products}).

\vspace{3mm}
\no
{\itshape
(a) {\bfseries 
      (100000)$\,\otimes\,$(000001) $=$ (100001)$\,\oplus\,$(000100)$\,\oplus\,$(100000)
    }
}

             At level 4 of the 1728-dimensional {\bf (100001)} irrep
             we find four states with weight $(000100)$. This weight space,
             however, is five-dimensional, and the computation of the
             state orthogonal to the previous four yields the highest
             weight state of the ${\bf 351}$ irrep. 
             (See table \ref{t:78cgc000100}.)
             Similarly, at level 11 we find 16-fold degenerate weight $(100000)$,
             while this reducible subspace unfolds to be 22-dimensional. Since there are
             five distinct states of the same weight in the ${\bf 351}$, there is
             room for one extra state. Once computed as orthogonal to all the
             other 21 states it becomes the highest weight state of the
             fundamental 27-dimensional {\bf (100000)} irrep. 
             Note that in table \ref{t:78cgc100000} we keep the labeling 
             of the five $(100000)$ states of the ${\bf 351}$ 
             consistent with the notation introduced in equations 
             (\ref{eq:100000in351P}) and (\ref{eq:100000in351}).

\vspace{3mm}
\no
{\itshape
(b) {\bfseries 
      (100000)$\,\otimes\,$(000100) $=$ (100100)$\,\oplus\,$(000011)$\,\oplus\,$(010000)$\,\oplus\,$(000010)
    }
}

             Lowering down to level 4 of the 7371-dimensional {\bf (100100)} irrep       
             we obtain four distinct $(000011)$ weight states spanned over 
             a five-dimensional reducible subspace. The last basis state in this subspace,
             orthogonal to the four from the ${\bf \ol{7371}}$, becomes the highest
             weight state of the ${\bf \ol{1728}}$. 
             (See table \ref{t:351cgc000011}.)
             This is a conjugate irrep to the ${\bf 1728}$ described in {\it (a)}.
             The lowering paths to the dominant weights in its weight system can be
             obtained from table \ref{t:paths_1728} (replacing 1 and 2 with 5 and 4, 
             and {\it vice versa}. Proceeding to level 7 
             a five-fold degenerate $(200000)$ dominant weight is found:
             \be
                 |200000_a\rangle = |100000\rangle\,|100000_a\rangle, 
                 \quad\quad a=2,\ldots 6
             \label{eq:200000b}
             \ee
             which, obviously, does not leave any extra space 
             for states outside of the ${\bf \ol{7371}}$. This is consistent
             with no observation in {\it (a)} of a dominant weight $(000020)$
             in the weight system of the ${\bf 1728}$, and also with the fact that
             there is no ${\bf (200000)}$ irrep on the right side of 
             (\ref{eq:weights_in_products}b). The charge conjugation operators 
             can be used to show that a five-fold degenerate weight subspace
             with CGCs equal to 1
             is then present at odd levels of the ${\bf \ol{7371}}$ from this level down
             until 
             subspace $(0000\ol{2}0)$ emerges at level 39.
             Next, at level 8 fifteen linearly independent $(010000)$ weight states
             are present, while the weight subspace turns out to be 20-dimensional.
             Not surprisingly there are four states which belong to the 
             weight system of the ${\bf \ol{1728}}$ (compare with {\it (a)} above),
             and the remaining basis state, orthogonal to the previous nineteen,
             represents the highest weight state of the ${\bf \ol{351}}$.
             Lastly, at level 15 we get the reducible $(000010)$ weight subspace,
             which is 66-dimensional. That makes room for the highest weight of the
             ${\bf \ol{27}}$, since there are 44 basis states present 
             in the ${\bf \ol{7371}}$
             together with 16  states of the ${\bf \ol{1728}}$. Additional five states
             of the ${\bf \ol{351}}$ should be expected based on 
             eqs.(\ref{eq:100000in351P}, \ref{eq:100000in351}).
             The CGCs for this subspace are presented in 
             tables \ref{t:351cgc000010} and  \ref{t:351cgc000010part2}.

\vspace{3mm}
\no
{\itshape
(c) {\bfseries 
       (100000)$\,\otimes\,$(000020) $=$ (100020)$\,\oplus\,$(000011)$\,\oplus\,$(000010)
    }
}

             In the construction of the 7722-dimensional ${\bf (100020)}$ irrep one finds 
             a dominant weight already at level 1, 
             \be 
                 |100100\rangle = |100000\rangle\,|000100\rangle.
             \ee
             It occupies a one-dimensional subspace, which
             is consistent with the absence of the ${\bf (100100)}$ irrep
             in product (\ref{eq:weights_in_products}c). 
             The first degenerate dominant weight is obtained at level 5. 
             There, a six-dimensional $(000011)$ weight subspace contains five
             linearly independent states of the ${\bf \ol{7722}}$. The basis in
             this reducible subspace is completed by the highest weight state 
             of the ${\bf \ol{1728}}$ (table \ref{t:351primecgc000011}).
             Proceeding further, there is no room
             for the highest weight state of a new irrep when dominant
             weights $(200000)$ and $(010000)$ are encountered at levels 8 and 9,
             respectively. The $(200000)$ subspace is four-dimensional and its basis 
             can be specified as in eq.(\ref{eq:200000b}). (The states are 
             now numbered as $a=3,4,5,6$.) The CGC decomposition of the
             $(010000)$ subspace can be found in table \ref{t:351primecgc010000}.
             Finally, at level 16 the last dominant weight in this product is
             unveiled. The reducible $(000010)$ weight subspace turns out to 
             be 57-dimensional,
             with 40 basis states coming from the ${\bf \ol{7722}}$ and 16 states 
             from the ${\bf \ol{1728}}$. The remaining state, orthogonal to them,
             becomes the highest weight state of the ${\bf \ol{27}}$ 
             (see tables \ref{t:351primecgc000010} and \ref{t:351primecgc000010part2}).

\vspace{3mm}
\no
{\itshape
(d) {\bfseries 
       (100000)$\,\otimes\,$(010000) $=$ (110000)$\,\oplus\,$(001000)$\,\oplus\,$(100010)$\,\oplus\,$(000001)
    }
}

             In this product, we find the dominant weight states encountered
             already in the decomposition of ${\bf 78}\otimes {\bf 78}$ and 
             ${\bf 27}\otimes {\bf \ol{27}}$. At level 2 of the
             $\bf (110000)$ weight system ({\it i.e.}, the ${\bf \ol{5824}}$ irrep) 
             we reach the 3-dimensional $(001000)$ subspace, with two states in the ${\bf \ol{5824}}$ 
             and the third one being the highest weight state of the ${\bf 2925}$,
             as shown in table  \ref{t:351barcgc001000}.
             Then following the lowering paths in table \ref{t:paths_5824bar}, 
             table II in Ref.\cite{abII}, and table I in Ref.\cite{abI} the dominant
             weights $(100010)$, $(000001)$,  and $(000000)$ follow at levels
             7, 12, and 23, respectively. The CGCs for these dominant weights
             can be found in tables \ref{t:351barcgc100010}--\ref{t:351barcgc000000part4}.
             The reducible $(000000)$ subspace is 135-dimensional and represents
             the most (technically) involved computation in this study. Obviously,
             it cannot (and does not) leave any room for the singlet since the two
             representations in the product are not conjugate to each other.

\vspace{3mm}
\no
{\itshape
(e) {\bfseries 
      (100000)$\,\otimes\,$(200000) $=$ (300000)$\,\oplus\,$(110000)$\,\oplus\,$(100010)
    }
}

             The 3003-dimensional ${\bf (300000)}$ irrep contains a dominant weight already at level 1:
             \be
                 |110000\rangle = (\,|\ol{1}10000\rangle\,|200000\rangle
                                  + \sqrt{2}\,|100000\rangle\,|010000\rangle)/\sqrt{3}.
             \ee
             The orthogonal combination
             \be
                 |110000\rangle = (\,\sqrt{2} |\ol{1}10000\rangle\,|200000\rangle
                                  -           |100000\rangle\,|010000\rangle)/\sqrt{3}
             \ee
             forms the highest weight state of the ${\bf \ol{5824}}$.
             Since then, the same dominant weights occur as in the weight system of the
             ${\bf \ol{5824}}$ described under {\it (d)} above. There are, however, no ${\bf 2925}$
             and ${\bf 78}$ irreps in this product (see tables \ref{t:351prime_barcgc001000},
             \ref{t:351prime_barcgc000001}, and \ref{t:351prime_barcgc000001part2}), 
             just the highest weight state of the ${\bf 650}$ completes the 13-dimensional $(100010)$ subspace 
             at level 8 (table \ref{t:351prime_barcgc100010}). The reducible $(000000)$ weight subspace 
             is 108-dimensional and its decomposition can be found 
             in tables \ref{t:351prime_barcgc000000part1}--\ref{t:351prime_barcgc000000part4}.

\section{Summary}

We have presented the Clebsch-Gordan decomposition of the $E_6$ tensor
products of the fundamental ${\bf 27}$ irrep 
with the 78- and 351-dimensional irreps. Analogous products involving
the ${\bf \ol{27}}$ instead of the ${\bf 27}$ can now be obtained trivially
by charge conjugation. It is straightforward to apply these results to
the construction of higher dimension operators in $E_6$ model building
\cite{ab_models}.

\section*{Appendix: The problem of Degenerate Weights}

\appendix

\def\sp#1#2{\langle\, #1 , #2\, \rangle}

Rules (\ref{eq:dfnN},\ref{eq:N-}) are insufficient for representations 
with degenerate weights at successive levels. 
For degenerate weights we must first identify a particular 
basis. Label  the degenerate basis states of weights
$(w+\alpha)$, $(w)$, and $(w-\alpha)$ as
\begin{eqnarray*}
&\mid&\!\! (w+\alpha)_{\Gamma} \rangle, \quad\mbox{where}\quad \Gamma = 1,\ldots D_{w+\alpha}, \\
&\mid&\!\! \;\; w_{c}\qquad\:  \rangle, \quad\mbox{where}\quad      c = 1,\ldots D_{w}, \qquad \mbox{and} \\
&\mid&\!\! (w-\alpha)_{C}      \rangle, \quad\mbox{where}\quad      C = 1,\ldots D_{w-\alpha}. 
\end{eqnarray*}
$D_w$ stands for the degeneracy of $(w)$.  
The basis states are in general non-orthogonal.
In our notation, they are always normalized to unity:
$\langle w_{c} \mid w_{c} \rangle = 1$.
The identity operator in the degenerate subspace is
\begin{eqnarray}
I\; &=& G_{ab} \mid w_a\, \rangle \langle\, w_b \mid, 
\cr
G_{ab} &=& (M^{-1})_{ab}, \quad \mbox{where}\quad M_{ab} = \langle\, w_a \mid w_b\, \rangle.
\end{eqnarray}

Although the basis is non-orthogonal, we can construct state
vectors which are orthogonal to any state except the state
we are interested in
\begin{eqnarray}
\mid      \hat{w_b}\, \rangle &=&  \mid w_a\, \rangle\: G_{ab}, \cr
\langle \,\hat{w_a}   \mid    &=&  G_{ab}  \: \langle\, w_b \mid,
\end{eqnarray}
which satisfy
\begin{equation}
\langle \,w_c \mid \hat{w_a}\, \rangle = \langle \,\hat{w_a} \mid w_c\, \rangle = \delta_{ac}.
\end{equation}
\noindent
A general raising or lowering of a degenerate weight state can be written as:
\begin{eqnarray}
E_{\:\alpha_i} \mid w_c \rangle &=&
    N_{\:\alpha_i,\,w_c \rightarrow (w+\alpha_i)_\Gamma} \;\,
                               \mid (w+\alpha_i)_\Gamma \rangle \\
E_{ -\alpha_i} \mid w_c \rangle &=&
    N_{ -\alpha_i,\,w_c \rightarrow (w-\alpha_i)_C} 
                               \mid (w-\alpha_i)_C \rangle
\label{eq:general_lowering_relation} 
\end{eqnarray}
where there is a possible sum over the states on the right hand side 
[compare (\ref{eq:general_lowering_relation})  with (\ref{eq:dfnN})]. 
The lowering normalization constant can then be expressed only as a sum of matrix elements
$N_{-\alpha_i,\,w_{a} \rightarrow (w-\alpha_i)_A}
= G^{(w-\alpha_i)}_{AB}\,
\langle (w-\alpha_i)_B \mid E_{-\alpha_i} \mid w_{a} \rangle $.\cite{footnote_3} 
Using $E_{\alpha} = E_{-\alpha}^{\dag}$
and the defining relation (\ref{eq:general_lowering_relation}) 
we derive
\begin{eqnarray}
& &
    N_{-\alpha,\, w_a \rightarrow (w-\alpha)_A}\, 
N^{*}_{-\alpha,\, w_b \rightarrow (w-\alpha)_B}\,
\langle \,(w-\alpha)_B \mid (w-\alpha)_A \rangle 
\qquad\qquad\qquad\qquad\qquad\qquad\qquad\qquad\qquad
\qquad\qquad\qquad\qquad\qquad\qquad\qquad\qquad\qquad
\cr
& & \qquad
= 
\langle w_b \mid E_{\alpha} E_{-\alpha} \mid w_a \rangle
\cr
& & \qquad
= 
\langle w_b \mid [ E_{\alpha}, E_{-\alpha} ] 
                +  E_{-\alpha} E_{\alpha} 
  \mid  w_a \rangle 
\cr
& & \qquad
= 
\langle w_b \mid w_a \rangle \,\sp{\alpha}{w} + \,
G^{w+\alpha}_{\Gamma\Delta} \,
\langle w_b \mid E_{-\alpha} \mid (w+\alpha)_\Gamma \rangle\,
\langle (w+\alpha)_\Delta \mid E_{\alpha} \mid  w_a \rangle 
\cr
& & \qquad
= 
\langle w_b \mid w_a \rangle \,\sp{\alpha}{w} + \,
G^{w+\alpha}_{\Gamma\Delta} \,
N_{-\alpha,\,(w+\alpha)_\Gamma \rightarrow w_c}
\langle w_b \mid w_c \rangle\,
N^{*}_{-\alpha,\, (w+\alpha)_\Delta \rightarrow w_d}
\langle w_d \mid w_a \rangle.
\end{eqnarray}

\noindent
Hence 
\begin{eqnarray}
  \mbox{\rule[-0.40cm]{0mm}{0.5cm}}
  (G^{w-\alpha})^{-1}_{AB}&\!\!\!
      N_{-\alpha,\, w_a \rightarrow (w-\alpha)_A} \,
  N^{*}_{-\alpha,\, w_b \rightarrow (w-\alpha)_B} 
  \qquad\qquad\qquad\qquad\qquad\qquad\qquad\qquad\qquad
 \cr
 =
 &\!
  (G^w)^{-1}_{ab} \sp{\alpha}{w} +\, 
  G^{w+\alpha}_{\Gamma\Delta} \,(G^w)^{-1}_{bc}(G^w)^{-1}_{da}\,
      N_{-\alpha,\, (w+\alpha)_\Gamma \rightarrow w_c}\,
  N^{*}_{-\alpha,\, (w+\alpha)_\Delta \rightarrow w_d}\,.
\label{eq:app2}
\end{eqnarray}
For $a=b$ we get:
\begin{eqnarray}
  \mbox{\rule[-0.40cm]{0mm}{0.5cm}}
  (G^{w-\alpha})^{-1}_{AB} &\!\!\!
      N_{-\alpha,\, w_a \rightarrow (w-\alpha)_A} \,
  N^{*}_{-\alpha,\, w_a \rightarrow (w-\alpha)_B} 
  \qquad\qquad\qquad\qquad\qquad\qquad\qquad\qquad\qquad
 \cr
 =
 &\!
  \sp{\alpha}{w} +\, 
  G^{w+\alpha}_{\Gamma\Delta} \,(G^w)^{-1}_{ac}(G^w)^{-1}_{ad}\,
      N_{-\alpha,\, (w+\alpha)_\Gamma \rightarrow w_c}\,
  N^{*}_{-\alpha,\, (w+\alpha)_\Delta \rightarrow w_d}\,.
  \qquad\qquad
\label{eq:app3}
\end{eqnarray}

\noindent
Another useful expression can be found
by contracting relation (\ref{eq:app2}) with $G^{w}_{ab}\:$:
\begin{eqnarray}
\mbox{\rule[-0.40cm]{0mm}{0.5cm}}
(G^w)_{ab} \!\!\!\!\!\! & 
(G^{w-\alpha})^{-1}_{AB}\;
    N_{-\alpha,\, w_a \rightarrow (w-\alpha)_A} \,
N^{*}_{-\alpha,\, w_b \rightarrow (w-\alpha)_B} 
\qquad\qquad\qquad\qquad\qquad\qquad\qquad\qquad
\cr
=
&\!
\sp{\alpha}{w} \, D_w + \,
G^{w+\alpha}_{\Gamma\Delta}\, (G^w)^{-1}_{cd} \,
    N_{-\alpha,\, (w+\alpha)_\Gamma \rightarrow w_c} \,
N^{*}_{-\alpha,\, (w-\alpha)_\Delta \rightarrow w_d} \,.
\qquad\qquad\qquad\qquad\qquad
\end{eqnarray}
This expression is easily iterated along a sequence
of lowerings with the same ladder operator:
\begin{eqnarray}
\mbox{\rule[-0.40cm]{0mm}{0.5cm}}
(G^w)_{ab} \!\!\!\!\!\! & 
(G^{w-\alpha})^{-1}_{AB}\;  
    N_{-\alpha,\, w_a \rightarrow (w-\alpha)_A} \, 
N^{*}_{-\alpha,\, w_b \rightarrow (w-\alpha)_B} 
\qquad\qquad\qquad\qquad\qquad\qquad\qquad\qquad\qquad\qquad
\cr
=
&\!
\sp{\alpha}{w} \, D_w + \sp{\alpha}{w+\alpha} \, D_{w+\alpha} +\,
G^{w+2\alpha}_{\gamma\delta}\, (G^{w+\alpha})^{-1}_{\Gamma\Delta} \,
    N_{-\alpha,\, (w+2\alpha)_\gamma \rightarrow (w+\alpha)_\Gamma}\,
N^{*}_{-\alpha,\, (w+2\alpha)_\delta \rightarrow (w+\alpha)_\Delta}
\cr
=
&\!
\sp{\alpha}{w}\,D_w  +          \sp{\alpha}{w+ \alpha}\, D_{w+ \alpha}
                     + \ldots + \sp{\alpha}{w+k\alpha}\, D_{w+k\alpha}\, , 
\qquad\qquad\qquad\qquad\qquad\qquad\quad
\end{eqnarray}
where $(w+k\alpha)$ is the highest weight in the $SU(2)$ subgroup
chain $(w)$, $(w+\alpha)$, $(w+2\alpha)$, $\ldots$ present in the weight system.

Finally, for completeness, when raising operators are applied, eq.(\ref{eq:app3})
can be written as
\begin{eqnarray}
  \mbox{\rule[-0.40cm]{0mm}{0.5cm}}
  (G^{w+\alpha})^{-1}_{\Gamma\Delta}&\!\!\!
      N_{\alpha,\, w_a \rightarrow (w+\alpha)_\Gamma} \,
  N^{*}_{\alpha,\, w_a \rightarrow (w+\alpha)_\Delta} 
  \qquad\qquad\qquad\qquad\qquad\qquad\qquad\qquad\qquad
 \cr
 =
 &\!
  - \sp{\alpha}{w} +\, 
  G^{w-\alpha}_{CD} \,(G^w)^{-1}_{ac}(G^w)^{-1}_{ad}\,
      N_{\alpha,\, (w-\alpha)_C \rightarrow w_c}\,
  N^{*}_{\alpha,\, (w-\alpha)_D \rightarrow w_d}.
  \qquad\qquad
\end{eqnarray}

\subsubsection*{Special Cases: Lowering within basis states}
 
Consider a series of states connected by repeated application
of the same lowering operator $E_{-\alpha}$. Choose any 
states with degenerate weights obtained in this series as part 
of the basis for the degenerate weights and label these states by $i$.
For the sequence: 
$\ldots, (w + \alpha)_i, w_i, (w-\alpha)_i, \ldots$ the
generalized recursion relation (\ref{eq:app3}) reduces to:
\be
  \mbox{\rule[-0.40cm]{0mm}{0.5cm}}
  \mid N_{-\alpha,\, w_i \rightarrow (w -\alpha)_i} \mid^2 
  \, = \, \sp{\alpha}{w} + \:
  G^{w+\alpha}_{\Gamma\Delta} (G^w)^{-1}_{ic}
                              (G^w)^{-1}_{id}
      N_{-\alpha,\, (w +\alpha)_\Gamma \rightarrow w_c}
  N^{*}_{-\alpha,\, (w +\alpha)_\Delta \rightarrow w_d}\, .
\ee
When $(w + \alpha)_i$ is the only state which can be lowered by $E_{-\alpha}$ 
to obtain a state of weight $w$ we get 
\be
  \mbox{\rule[-0.40cm]{0mm}{0.5cm}}
  \mid N_{-\alpha,\, w_i \rightarrow (w -\alpha)_i} \mid^2 
  = \sp{\alpha}{w} + \:
  G^{(w+\alpha)}_{ii} 
  |N_{-\alpha,\, (w +\alpha)_i \rightarrow w_i}|^2.
\ee
This includes the case of a non-degenerate $(w + \alpha)$ weight subspace.
When $(w + \alpha)$ is non-degenerate $G^{(w+\alpha)}_{ii} = 1$, which
further simplifies the above relation.

A special case of interest is lowering the degenerate
zero weight states of the adjoint representation which
correspond to the Cartan sub-algebra.  These
degenerate weight states can be labeled $\mid (0)_i\, \rangle$ where 
the $i$-th degenerate weight is obtained by 
$E_{-\alpha_i} \mid \alpha_i\, \rangle \propto\:
\mid (0)_i\, \rangle$. This basis, however, is not orthogonal.
When lowering such a basis state the general formula 
(\ref{eq:app3}) reduces to:
\be
    \mid N_{-\alpha_i, (0)_j \rightarrow (-\alpha_i)} \mid^2
 \: = \:
    \left[ (G^{(0)})^{-1}_{ji} \right]^2
    \mid N_{-\alpha_i,\, (\alpha_i) \rightarrow (0)_i }\mid^2 
 \: = \: 
    \langle (0)_j\, |\, (0)_i\rangle^2\;\; 
    \mid N_{-\alpha_i,\, (\alpha_i) \rightarrow (0)_i }\mid^2 .
\ee
This result 
is consistent with formula (\ref{eq:N-}) in section \ref{sc:MathPrelim}
when applied to the zero weight states of the ${\bf 78}$ in $E_6$.

\newpage

\protect
\begin{table}
\caption{ 
{\bf Bases in the dominant weight subspaces 
of the 1728-dimensional $(100001)$ irrep.} 
}
\label{t:paths_1728}
\begin{tabular}{|l|c||l|c|}
\hline
                                   &                &                                   &              \\
  Weight state                     &  Lowering path &  Weight state                     & Lowering path\\
                                   &                &                                   &              \\
\hline 						    						       
                                   &                &                                   &              \\
$\;\;\mid\!\! 000100_6     \rangle$&  6321          &$\;\;\mid\!\! 100000_{\: 6}\rangle$& 65324436321  \\
$\;\;\mid\!\! 000100_3     \rangle$&  3621          &$\;\;\mid\!\! 100000_{\: 7}\rangle$& 64534236321  \\
$\;\;\mid\!\! 000100_2     \rangle$&  2361          &$\;\;\mid\!\! 100000_{\: 8}\rangle$& 63214534236  \\
$\;\;\mid\!\! 000100_1     \rangle$&  1236          &$\;\;\mid\!\! 100000_{\: 9}\rangle$& 63243654321  \\
                                   &                &$\;\;\mid\!\! 100000_{10}  \rangle$& 53624436321  \\
 \cline{1-2}                       &                &$\;\;\mid\!\! 100000_{11}  \rangle$& 54321634236  \\
$\;\;\mid\!\! 100000_{\: 1}\rangle$& 12364534236    &$\;\;\mid\!\! 100000_{12}  \rangle$& 32164534236  \\
$\;\;\mid\!\! 100000_{\: 2}\rangle$& 21364534236    &$\;\;\mid\!\! 100000_{13}  \rangle$& 32643654321  \\
$\;\;\mid\!\! 100000_{\: 3}\rangle$& 25364436321    &$\;\;\mid\!\! 100000_{14}  \rangle$& 32643254361  \\
$\;\;\mid\!\! 100000_{\: 4}\rangle$& 24534636321    &$\;\;\mid\!\! 100000_{15}  \rangle$& 45346236321  \\
$\;\;\mid\!\! 100000_{\: 5}\rangle$& 23643254361    &$\;\;\mid\!\! 100000_{16}  \rangle$& 45321634236  \\
                                   &                &                                   &              \\
\hline					                    
\end{tabular}				                    
\end{table}

\protect
\begin{table}
\caption{ 
{\bf Bases in the dominant weight subspaces 
of the $(100100)$ irrep, the ${\bf \ol{7371}}$.} 
$\mid\!\! 000010_n \rangle$ states are marked $\mid\!\! \ol{F}_n \rangle$ for brevity.
}
\label{t:paths_7371bar}
\small
\begin{tabular}{|l|c||l|c||l|c|}
\hline
                            &              & \multicolumn{4}{c|}{ }                                                    \\
 Weight state               & Lowering path& \multicolumn{4}{c|}{Lowering paths to $(000010)$ weight states }          \\
                            &              & \multicolumn{4}{c|}{ }                                                    \\
\hline 
                            &              &                             &                  &                             &                  \\
$\gapA 000011_4     \rangle$&  4321        & $\gapA \ol{F}_{\: 1}\rangle$& 514362236434321  & $\gapA \ol{F}_{23}  \rangle$& 645342136234321  \\
$\gapA 000011_3     \rangle$&  3421        & $\gapA \ol{F}_{\: 2}\rangle$& 563214436234321  & $\gapA \ol{F}_{24}  \rangle$& 643452136234321  \\
$\gapA 000011_2     \rangle$&  2341        & $\gapA \ol{F}_{\: 3}\rangle$& 536214436234321  & $\gapA \ol{F}_{25}  \rangle$& 643621345234321  \\
$\gapA 000011_1     \rangle$&  1234        & $\gapA \ol{F}_{\: 4}\rangle$& 523614436234321  & $\gapA \ol{F}_{26}  \rangle$& 636231245434321  \\
\end{tabular}				                    
\end{table}				                    
\normalsize

\protect
\begin{table}
\caption{ 
{\bf Bases in the dominant weight subspaces 
of the $(100020)$ irrep, the ${\bf \ol{7722}}$.} 
(100100) weight is left out as trivial.
$\mid\!\! 000010_n \rangle$ states are marked $\mid\!\! \ol{F}_n \rangle$ for brevity.
}
\label{t:paths_7722bar}
\small
\begin{tabular}{|l|c||l|c||l|c|}
\hline
                            &              & \multicolumn{4}{c|}{ }                                                    \\
 Weight state               & Lowering path& \multicolumn{4}{c|}{Lowering paths to $(000010)$ weight states }          \\
                            &              & \multicolumn{4}{c|}{ }                                                    \\
\hline 
                            &              &                             &                  &                             &                  \\
$\gapA 000011_5     \rangle$& 54321        & $\gapA \ol{F}_{\: 1}\rangle$& 5632144362345321 & $\gapA \ol{F}_{21}  \rangle$& 3164213623452345 \\
$\gapA 000011_4     \rangle$& 45321        & $\gapA \ol{F}_{\: 2}\rangle$& 5362144362345321 & $\gapA \ol{F}_{22}  \rangle$& 3164223645345321 \\
$\gapA 000011_3     \rangle$& 34521        & $\gapA \ol{F}_{\: 3}\rangle$& 5236144362345321 & $\gapA \ol{F}_{23}  \rangle$& 3445231266345321 \\
$\gapA 000011_2     \rangle$& 23451        & $\gapA \ol{F}_{\: 4}\rangle$& 5123644362345321 & $\gapA \ol{F}_{24}  \rangle$& 3452341266345321 \\
\end{tabular}				                    
\end{table}				                    
\normalsize

\protect
\begin{table}
\caption{ 
{\bf Bases in the dominant weight subspaces 
of the $(110000)$ irrep, the ${\bf \ol{5824}}$.} 
}
\label{t:paths_5824bar}
\small
\begin{tabular}{|l|c||l|c||l|c|}
\hline
                            &              & \multicolumn{4}{c|}{ }                                              \\
 Weight state               & Lowering path& \multicolumn{4}{c|}{Lowering paths to zero weight states }          \\
                            &              & \multicolumn{4}{c|}{ }                                              \\
\hline 
                            &              &                        &                         &                        &                         \\
$\gapA 001000_2     \rangle$&  21          & $\gapA 0_{\: 1}\rangle$& 65241364345236342133221 & $\gapA 0_{33}  \rangle$& 51453643212233664433221 \\
$\gapA 001000_1     \rangle$&  12          & $\gapA 0_{\: 2}\rangle$& 65241363344521322364321 & $\gapA 0_{34}  \rangle$& 51436213452233664433221 \\
                            &              & $\gapA 0_{\: 3}\rangle$& 65142364345236342133221 & $\gapA 0_{35}  \rangle$& 51436213422334566321432 \\
 \cline{1-2}                &              & $\gapA 0_{\: 4}\rangle$& 65142363344521322364321 & $\gapA 0_{36}  \rangle$& 53623124536436214433221 \\
\end{tabular}				                    
\end{table}				                    
\normalsize

\protect
\begin{table}
\caption{ 
{\bf Bases in the dominant weight subspaces 
of the 3003-dimensional $(300000)$ irrep.} 
(110000) weight is left out as trivial.
}
\label{t:paths_3003}
\begin{tabular}{|l|c||l|c|}
\hline
                                   &                 &                                   &                          \\
  Weight state                     &  Lowering path  &  Weight state                     &  Lowering path           \\
                                   &                 &                                   &                          \\
\hline 						    
                                   &                 &                                   &                          \\
$\;\;\mid\!\! 001000       \rangle$&  211            &$\;\;\mid\!\! 000000_{\: 1}\rangle$& 652413643453323643222111 \\
                                   &                 &$\;\;\mid\!\! 000000_{\: 2}\rangle$& 651423643453323643222111 \\
 \cline{1-2}                       &                 &$\;\;\mid\!\! 000000_{\: 3}\rangle$& 653623124536444333222111 \\
$\;\;\mid\!\! 100010_6     \rangle$&  64332211       &$\;\;\mid\!\! 000000_{\: 4}\rangle$& 653214436323145236432211 \\
\hline					                    
\end{tabular}				                    
\end{table}


\protect
\begin{table}
\caption{ 
{\bf CG coefficients for (000100) dominant weight in (100000)$\otimes$(000001).}
Each entry should be divided by the respective number in the last row to keep
the states normalized to 1.
}
\label{t:78cgc000100}
\footnotesize
\begin{tabular}{|@{\hspace{0.5mm}}c
                |c@{$\,$}c@{$\,$}c@{$\,$}c@{$\,$}
                |c@{$\,$}|}
\hline
\mbox{ } & \multicolumn{4}{|c|}{\mbox{ }}                  & {\mbox{ }}                  \\
\mbox{ } & \multicolumn{4}{ c|}{\normalsize \em (100001) } & {\normalsize \em (000100) } \\
\mbox{ } & \multicolumn{4}{|c|}{\mbox{ }}                  & {\mbox{ }}                  \\
\cline{2-6}
 & 
   $\,\mid$000100$_{6}\rangle$ & 
   $\,\mid$000100$_{3}\rangle$ & 
   $\,\mid$000100$_{2}\rangle$ & 
   $\,\mid$000100$_{1}\rangle$ & 
   $\,\mid$000100$        \rangle$ \\ 
\hline
\AA00010\1\MM000001\ZZ    &  1  &      &      &      & $\;\,$1  \\
\AA00\1101\MM00100\1\ZZ   &  1  &   1  &      &      &      -1  \\
\AA0\11000\MM01\1100\ZZ   &     &   1  &   1  &      & $\;\,$1  \\
\AA\110000\MM1\10100\ZZ   &     &      &   1  &   1  &      -1  \\
\AA100000\MM\100100\ZZ    &     &      &      &   1  & $\;\,$1  \\
\hline
                          & $\sqrt{2}$ & $\sqrt{2}$ & $\sqrt{2}$ & $\sqrt{2}$ &$\sqrt{5}$ \\
\hline
\end{tabular}
\end{table}

\protect
\begin{table}
\caption{ 
{\bf CG coefficients for (100000) dominant weight in (100000)$\otimes$(000001).}
The fundamental $(100000)$ irrep is marked as $F$, and its highest weight state as \AA F\ZZ\.
\AA n\ZZ\ is an abbreviation for \AA100000$_n$\ZZ. Numbering of the degenerate states is 
consistent with table \ref{t:paths_1728} and eqs.(\ref{eq:100000in351P}--\ref{eq:100000in351}).
Each CGC should be divided by the respective number in the last row to maintain
$\langle n\!\mid\! n\rangle = 1$.
}
\label{t:78cgc100000}
\footnotesize

\end{table}


\protect
\begin{table}
\caption{ 
{\bf CG coefficients for (000011) dominant weight in (100000)$\otimes$(000100).}
Each entry should be divided by the respective number in the last row to keep
the states normalized to 1.
}
\label{t:351cgc000011}
\footnotesize
\begin{tabular}{|@{\hspace{0.5mm}}c
                |c@{$\,$}c@{$\,$}c@{$\,$}c@{$\,$}
                |c@{$\,$}|}
\hline
\mbox{ } & \multicolumn{4}{|c|}{\mbox{ }}                  & {\mbox{ }}                  \\
\mbox{ } & \multicolumn{4}{ c|}{\normalsize \em (100100) } & {\normalsize \em (000011) } \\
\mbox{ } & \multicolumn{4}{|c|}{\mbox{ }}                  & {\mbox{ }}                  \\
\cline{2-6}
 & 
   $\,\mid$000011$_{4}\rangle$ & 
   $\,\mid$000011$_{3}\rangle$ & 
   $\,\mid$000011$_{2}\rangle$ & 
   $\,\mid$000011$_{1}\rangle$ & 
   $\,\mid$000011$        \rangle$ \\ 
\hline
\AA000\111\MM000100\ZZ    &  1  &      &      &      & $\;\,$1  \\
\AA00\1101\MM001\110\ZZ   &  1  &   1  &      &      &      -1  \\
\AA0\11000\MM01\1011\ZZ   &     &   1  &   1  &      & $\;\,$1  \\
\AA\110000\MM1\10011\ZZ   &     &      &   1  &   1  &      -1  \\
\AA100000\MM\100011\ZZ    &     &      &      &   1  & $\;\,$1  \\
\hline
                          & $\sqrt{2}$ & $\sqrt{2}$ & $\sqrt{2}$ & $\sqrt{2}$ &$\sqrt{5}$ \\
\hline
\end{tabular}
\end{table}

\protect
\begin{table}
\caption{ 
{\bf CG coefficients for (010000) dominant weight in (100000)$\otimes$(000100).}
The $(010000)$ irrep is marked as $\ol{G}$, and its highest weight state as $\mid \ol{G}\rangle$.
\AA n\ZZ\ is an abbreviation for \AA010000$_n$\ZZ. Numbering of the degenerate
states is consistent with tables \ref{t:paths_7371bar} and \ref{t:paths_1728}.
Each CGC should be divided by the respective number in the last row to maintain
$\langle n\!\mid\! n\rangle = 1$.
}
\label{t:351cgc010000}
\footnotesize

\end{table}

\clearpage

\protect
\begin{table}
\caption{ 
{\bf CG coefficients for (000010) dominant weight in (100000)$\otimes$(000100).}
\AA n\ZZ\ is an abbreviation for \AA000010$_n$\ZZ. Numbering of the degenerate
states is consistent with tables \ref{t:paths_7371bar}, and \ref{t:paths_1728}.
Each CGC should be divided by the respective number in the last row to maintain
$\langle n\!\mid\! n\rangle = 1$.
}
\label{t:351cgc000010}
\tiny
\end{table}

\clearpage

\protect
\begin{table}
\caption{ 
{\bf CG coefficients for (000010) dominant weight in (100000)$\otimes$(000100).}
The fundamental $(100000)$ irrep is marked as $\ol{F}$, and its highest weight state as $\mid \ol{F}\rangle$.
\AA n\ZZ\ is an abbreviation for \AA000010$_n$\ZZ. Numbering of the degenerate
states is consistent with tables \ref{t:paths_7371bar}, \ref{t:paths_1728}, 
and eqs.(\ref{eq:100000in351P}--\ref{eq:100000in351}).
Each CGC should be divided by the respective number in the last row to maintain
$\langle n\!\mid\! n\rangle = 1$.
}
\label{t:351cgc000010part2}
\tiny

\end{table}

\clearpage


\protect
\begin{table}
\caption{ 
{\bf CG coefficients for (000011) dominant weight in (100000)$\otimes$(000020).}
%
Each entry should be divided by the respective number in the last row to keep
the states normalized to 1.
}
\label{t:351primecgc000011}
\footnotesize

\end{table}

\protect
\begin{table}
\caption{ 
{\bf CG coefficients for (010000) dominant weight in (100000)$\otimes$(000020).}
\AA n\ZZ\ is an abbreviation for \AA010000$_n$\ZZ. Numbering of the degenerate
states is consistent with tables \ref{t:paths_7722bar}, and \ref{t:paths_1728}.
Each CGC should be divided by the respective number in the last row to maintain
$\langle n\!\mid\! n\rangle = 1$.
}
\label{t:351primecgc010000}
\footnotesize
\end{table}

\clearpage

\protect
\begin{table}
\caption{ 
{\bf CG coefficients for (000010) dominant weight in (100000)$\otimes$(000020).}
\AA n\ZZ\ is an abbreviation for \AA000010$_n$\ZZ. Numbering of the degenerate
states is consistent with tables \ref{t:paths_7722bar}, and \ref{t:paths_1728}.
Each CGC should be divided by the respective number in the last row to maintain
$\langle n\!\mid\! n\rangle = 1$.
}
\label{t:351primecgc000010}
\tiny
\end{table}

\clearpage

\protect
\begin{table}
\caption{ 
{\bf CG coefficients for (000010) dominant weight in (100000)$\otimes$(000020).}
The fundamental $(100000)$ irrep is marked as $\ol{F}$, and its highest weight state as $\mid \ol{F}\rangle$.
\AA n\ZZ\ is an abbreviation for \AA000010$_n$\ZZ. Numbering of the degenerate
states is consistent with tables \ref{t:paths_7722bar}, and \ref{t:paths_1728}.
%
%
%
%
%
%
Each CGC should be divided by the respective number in the last row to maintain
$\langle n\!\mid\! n\rangle = 1$.
}
\label{t:351primecgc000010part2}
\tiny

\end{table}

\clearpage


\protect
\begin{table}
\caption{ 
{\bf CG coefficients for (001000) dominant weight in (100000)$\otimes$(010000).}
%
Each entry should be divided by the respective number in the last row to keep
the states normalized to 1.
}
\label{t:351barcgc001000}
\footnotesize

\end{table}

\protect
\begin{table}
\caption{ 
{\bf CG coefficients for (100010) dominant weight in (100000)$\otimes$(010000).}
\AA n\ZZ\ is an abbreviation for \AA100010$_n$\ZZ. Numbering of the degenerate
states is consistent with table \ref{t:paths_5824bar}, and table II in \cite{abII}.
Each CGC should be divided by the respective number in the last row to maintain
$\langle n\!\mid\! n\rangle = 1$.
}
\label{t:351barcgc100010}
\footnotesize
\end{table}

\clearpage

\protect
\begin{table}
\caption{ 
{\bf CG coefficients for (000001) dominant weight states of the 5824-dimensional 
(110000) irrep in (100000)$\otimes$(010000).}
\AA n\ZZ\ is an abbreviation for \AA000001$_n$\ZZ. Numbering of the degenerate
states is consistent with table \ref{t:paths_5824bar}.
Each CGC should be divided by the respective number in the last row to maintain
$\langle n\!\mid\! n\rangle = 1$.
}
\label{t:351barcgc000001}
\scriptsize
\end{table}

\clearpage

\protect
\begin{table}
\caption{ 
{\bf CG coefficients for (000001) dominant weight in (100000)$\otimes$(010000).}
The adjoint $(000001)$ irrep is marked as $A$, and its highest weight state as $\mid \!A\rangle$.
\AA n\ZZ\ is an abbreviation for \AA000001$_n$\ZZ. Numbering of the degenerate
states is consistent with table II in \cite{abII}, and table I in \cite{abI}.
Each CGC should be divided by the respective number in the last row to maintain
$\langle n\!\mid\! n\rangle = 1$.
}
\label{t:351barcgc000001part2}
\scriptsize

\end{table}

\clearpage

\protect
\begin{table}
\caption{ 
{\bf CG coefficients for the first 32 (000000) dominant weight states of the 5824-dimensional
(110000) irrep in the product (100000)$\otimes$(010000).} (The remaining 32 states of this
irrep with the same weight are shown in table \ref{t:351barcgc000000part2}.)
\AA n\ZZ\ is an abbreviation for \AA000000$_n$\ZZ. Numbering of the degenerate
states is consistent with table \ref{t:paths_5824bar}.
Each CGC should be divided by the respective number in the last row of the table to maintain
$\langle n\!\mid\! n\rangle = 1$.
}
\label{t:351barcgc000000part1}
\tiny

\end{table}

\clearpage

%
%

\protect
\begin{table}
\caption{ 
{\bf CG coefficients for the remaining 32 (000000) dominant weight states of the 5824-dimensional
(110000) irrep in the product (100000)$\otimes$(010000).} (The first 32 states of this
irrep with the same weight are shown in table \ref{t:351barcgc000000part1}.)
\AA n\ZZ\ is an abbreviation for \AA000000$_n$\ZZ. Numbering of the degenerate
states is consistent with table \ref{t:paths_5824bar}.
Each CGC should be divided by the respective number in the last row to maintain
$\langle n\!\mid\! n\rangle = 1$.
}
\label{t:351barcgc000000part2}
\tiny

\end{table}

\clearpage

%
%

\protect
\begin{table}
\caption{ 
{\bf CG coefficients for the first 36 (000000) dominant weight states of the 2925-dimensional
(001000) irrep in the product (100000)$\otimes$(010000).} (The remaining 9 states of this
irrep with the same weight are shown in table \ref{t:351barcgc000000part4}.)
\AA n\ZZ\ is an abbreviation for \AA000000$_n$\ZZ. Numbering of the degenerate
states is consistent with table II. in ref.\cite{abII}.
Each CGC should be divided by the respective number in the last row to maintain
$\langle n\!\mid\! n\rangle = 1$.
}
\label{t:351barcgc000000part3}
\tiny

\end{table}

\clearpage

%
%

\protect
\begin{table}
\caption{ 
{\bf CG coefficients for the (000000) dominant weight states of the 2925-dimensional
(001000) irrep, 650-dimensional (100010) irrep, and 78-dimensional (000001) irrep
in the product (100000)$\otimes$(010000).} 
\AA n\ZZ\ is an abbreviation for \AA000000$_n$\ZZ. Numbering of the degenerate
states is consistent with table II. in ref.\cite{abII} and table I in ref.\cite{abI}.
Each CGC should be divided by the respective number in the last row to maintain
$\langle n\!\mid\! n\rangle = 1$.
}
\label{t:351barcgc000000part4}  
\tiny

\end{table}

\clearpage

%
%


\protect
\begin{table}
\caption{ 
{\bf CG coefficients for (001000) dominant weight in (100000)$\otimes$(200000).}
%
Each entry should be divided by the respective number in the last row to keep
the states normalized to 1.
}
\label{t:351prime_barcgc001000}
\footnotesize

\end{table}

\protect
\begin{table}
\caption{ 
{\bf CG coefficients for (100010) dominant weight in (100000)$\otimes$(200000).}
\AA n\ZZ\ is an abbreviation for \AA100010$_n$\ZZ. Numbering of the degenerate
states is consistent with tables \ref{t:paths_3003} and \ref{t:paths_5824bar}.
Each CGC should be divided by the respective number in the last row to maintain
$\langle n\!\mid\! n\rangle = 1$.
}
\label{t:351prime_barcgc100010}
\footnotesize
\end{table}

\clearpage

\protect
\begin{table}
\caption{ 
{\bf CG coefficients for (000001) dominant weight states of the 3003-dimensional (300000) irrep
and 650-dimensional (100010) irrep in (100000)$\otimes$(200000).}
\AA n\ZZ\ is an abbreviation for \AA000001$_n$\ZZ. Numbering of the degenerate
states is consistent with table \ref{t:paths_3003}, and table I in \cite{abI}.
Each CGC should be divided by the respective number in the last row to maintain
$\langle n\!\mid\! n\rangle = 1$.
}
\label{t:351prime_barcgc000001}
\scriptsize

\end{table}

\clearpage

\protect
\begin{table}
\caption{ 
{\bf CG coefficients for (000001) dominant weight states of the 5824-dimensional 
(110000) irrep in (100000)$\otimes$(200000).}
\AA n\ZZ\ is an abbreviation for \AA000001$_n$\ZZ. Numbering of the degenerate
states is consistent with table \ref{t:paths_5824bar}.
Each CGC should be divided by the respective number in the last row to maintain
$\langle n\!\mid\! n\rangle = 1$.
}
\label{t:351prime_barcgc000001part2}
\scriptsize
\end{table}

\clearpage


\protect
\begin{table}
\caption{ 
{\bf CG coefficients for the (000000) dominant weight states of the 3003-dimensional
(300000) irrep in the product (100000)$\otimes$(200000).} 
\AA n\ZZ\ is an abbreviation for \AA000000$_n$\ZZ. Numbering of the degenerate
states is consistent with table \ref{t:paths_3003}.
Each CGC should be divided by the respective number in the last row of the table to maintain
$\langle n\!\mid\! n\rangle = 1$.
}
\label{t:351prime_barcgc000000part1}
\tiny

\end{table}

\clearpage

%
%

\protect
\begin{table}
\caption{ 
{\bf CG coefficients for the first 32 (000000) dominant weight states of the 5824-dimensional
(110000) irrep in the product (100000)$\otimes$(200000).} (The remaining 32 states of this
irrep with the same weight are shown in table \ref{t:351prime_barcgc000000part3}.)
\AA n\ZZ\ is an abbreviation for \AA000000$_n$\ZZ. Numbering of the degenerate
states is consistent with table \ref{t:paths_5824bar}.
Each CGC should be divided by the respective number in the last row of the table to maintain
$\langle n\!\mid\! n\rangle = 1$.
}
\label{t:351prime_barcgc000000part2}
\tiny

\end{table}

\clearpage

%
%

\protect
\begin{table}
\caption{ 
{\bf CG coefficients for the remaining 32 (000000) dominant weight states of the 5824-dimensional
(110000) irrep in the product (100000)$\otimes$(200000).} (The first 32 states of this
irrep with the same weight are shown in table \ref{t:351prime_barcgc000000part2}.)
\AA n\ZZ\ is an abbreviation for \AA000000$_n$\ZZ. Numbering of the degenerate
states is consistent with table \ref{t:paths_5824bar}.
Each CGC should be divided by the respective number in the last row to maintain
$\langle n\!\mid\! n\rangle = 1$.
}
\label{t:351prime_barcgc000000part3}
\tiny

\end{table}

\clearpage

%
%

\protect
\begin{table}
\caption{ 
{\bf CG coefficients for the (000000) dominant weight states of the 650-dimensional (100010) irrep
in the product (100000)$\otimes$(200000).} 
\AA n\ZZ\ is an abbreviation for \AA000000$_n$\ZZ. Numbering of the degenerate
states is consistent with table I in ref.\cite{abI}.
Each CGC should be divided by the respective number in the last row to maintain
$\langle n\!\mid\! n\rangle = 1$.
}
\label{t:351prime_barcgc000000part4}  
\tiny

\end{table}



\begin{thebibliography}{99}
\bibitem{footnote_1} For roots and weights we use the notation where $\ol{x}\equiv -x$.
\bibitem{footnote_2} Some applications of the charge conjugation operators 
                     can be found in ref.\cite{abI}.
\bibitem{footnote_3} It is not always possible to choose a single basis
                     for all of the states of degenerate weights in a representation 
                     such that lowering a particular state results
                     in a particluar state of lower weight (as opposed to a linear combination
                     of weights) for all possible lowerings.  The simplest example of 
                     this is the 27 dimensional representation of $SU(3)$.
\end{thebibliography}
\end{document}